\documentclass{article}
\usepackage{spconf,amsmath,graphicx,booktabs,array,multirow,siunitx,amssymb,subcaption,float,afterpage, longtable}
\usepackage[ruled,vlined]{algorithm2e}
\usepackage{adjustbox}
\usepackage{caption}
\usepackage{supertabular}
\usepackage{textcomp}
\usepackage{placeins}
\usepackage{needspace}
\usepackage{afterpage}

\title{Towards Biomarker Discovery for Early Cerebral Palsy Detection: Evaluating Explanations Through Kinematic Perturbations}

\name{Kimji N. Pellano$^{1}$, Inga Strümke$^{2}$, Daniel Groos$^{1}$, Lars Adde$^{3}$, Pål Haugen$^{2}$, Espen Alexander F. Ihlen$^{1}$} 
\address{
\begin{tabular}{@{}l@{}}
\small $^{1}$ Department of Neuromedicine and Movement Science, Faculty of Medicine and Health Sciences, \\
\small Norwegian University of Science and Technology, 7034 Trondheim, Norway\\
\small$^{2}$ Department of Computer Science, Faculty of Information Technology and Electrical Engineering, \\
\small Norwegian University of Science and Technology, 7034 Trondheim, Norway\\
\small$^{3}$ Department of Clinical and Molecular Medicine, Faculty of Medicine and Health Sciences,\\
\small Norwegian University of Science and Technology, 7034 Trondheim, Norway\\
\end{tabular}
}

\begin{document}
%
\maketitle
\begin{abstract}

Cerebral Palsy (CP) is a prevalent motor disability in children, for which early detection can significantly improve treatment outcomes. While skeleton-based Graph Convolutional Network (GCN) models have shown promise in automatically predicting CP risk from infant videos, their ``black-box” nature raises concerns about clinical explainability. To address this, we introduce a perturbation framework tailored for infant movement features and use it to compare two explainable AI (XAI) methods: Class Activation Mapping (CAM) and Gradient-weighted Class Activation Mapping (Grad-CAM). First, we identify significant and non-significant body keypoints in very low and very high risk infant video snippets based on the XAI attribution scores. We then conduct targeted velocity and angular perturbations, both individually and in combination, on these keypoints to assess how the GCN model’s risk predictions change. Our results indicate that velocity-driven features of the arms, hips, and legs have a dominant influence on CP risk predictions, while angular perturbations have a more modest impact. Furthermore, CAM and Grad-CAM show partial convergence in their explanations for both low and high CP risk groups. Our findings demonstrate the use of XAI-driven movement analysis for early CP prediction, and offer insights into potential movement-based biomarker discovery that warrant further clinical validation.

\end{abstract}
\section{Introduction}
\label{sec:introduction}

Cerebral Palsy (CP) is the most common motor disability in childhood, affecting approximately 2.11 per 1,000 live births worldwide \cite{oskoui2013update}. Early detection of CP is crucial for initiating timely interventions that can significantly improve outcomes and quality of life for affected individuals \cite{novak2017early}. Prechtl's General Movements Assessment (GMA) has emerged as a reliable method for early CP detection, which involves the systematic observation of spontaneous infant movements, particularly during the fidgety movement (FM) period at 3-5 months post-term age \cite{ricci2018feasibility}. However, the widespread implementation of GMA is challenging due to the need for trained experts and the subjective nature of the assessment \cite{silva2021future}.

Advancements in artificial intelligence (AI), particularly in computer vision and deep learning, have opened new possibilities for automated CP detection. Pose estimation algorithms can extract human skeletal data from videos \cite{dubey2023comprehensive, knap2024human}, which can then be processed to extract quantitative kinematic information. Deep learning models, especially those with Graph Convolutional Networks (GCNs), can then be used to analyze this spatio-temporal data from infant movements for CP prediction \cite{zhang2022cp, groos2022development, zhang2022cerebral}.

However, the ``black box" nature of deep learning models presents a significant challenge for their adoption in clinical settings. The lack of explainability raises concerns in medical applications where understanding the rationale behind a prediction is as important as the prediction itself \cite{teng2022survey, frasca2024explainable}. Various explainable AI (XAI) methods have been developed to address this challenge, such as Class Activation Mapping (CAM) \cite{zhou2016learning} and Gradient-weighted Class Activation Mapping (Grad-CAM) \cite{selvaraju2020grad}. While these methods have been successfully applied in various medical imaging applications \cite{schraut2023multi, t2024enhancing, xiao2021visualization}, their effectiveness in explaining predictions based on human movement data remains relatively unexplored. In our previous study \cite{pellano2024evaluating}, we conducted the initial work of evaluating CAM and Grad-CAM for a GCN-based CP-prediction model using established XAI metrics. However, while these XAI metrics provide valuable insights into the mathematical reliability of the explanations \cite{agarwal2022openxai, alvarez2018towards, zhou2021evaluating, markus2021role, alvarez2018robustness, agarwal2022rethinking}, they do not reveal how the identified significant and non-significant features relate to the underlying biomechanical features (e.g. joint angle or velocity changes) that can potentially help distinguish typical from atypical infant movements.

This study further extends our investigation into the explainability of CP prediction by analyzing motion perturbations through a previously developed ensemble GCN model. Using CAM and Grad-CAM feature attribution scores, we first identify the significant and non-significant body points that influence the model's risk predictions. We then systematically modify the velocity and angular motion of the said points. By examining how the resulting risk scores change in response to these biomechanically informed perturbations, we gain deeper insights into the underlying movement characteristics that contribute to CP risk assessment. The findings can provide a basis for future clinical validation to assess their relevance in CP diagnosis, and can be used to explain AI-based early CP prediction results.

This paper makes several contributions to the field:

\begin{itemize}
    \item Introduces a biomechanically-informed feature perturbation framework for XAI methods in movement analysis. This can be generalized to other applications such as gait analysis \cite{nguyen2016skeleton} and sports science \cite{guo2021attention}.

    \item Presents a comparison of CAM and Grad-CAM methods for explaining CP risk predictions in a GCN-based model.

    \item Provides insights into potential CP biomarker discovery by identifying consistently important joints across different risk groups and examining their underlying kinematic properties.

    \item Identifies the relative importance of kinematic features (velocity and angular motion) in CP risk assessment, which may guide future research into potential movement-based biomarkers for CP detection.

    \item Facilitates the clinical adoption of AI-based CP detection by demonstrating how XAI methods can provide explanations that are important for transparent and explainable decisions in pediatric care.
\end{itemize}

\subsection{CP Prediction Model}

\begin{figure*}[htbp]
    \centering
    \begin{subfigure}[t]{0.8\textwidth}
        \centering
        \includegraphics[width=\textwidth]{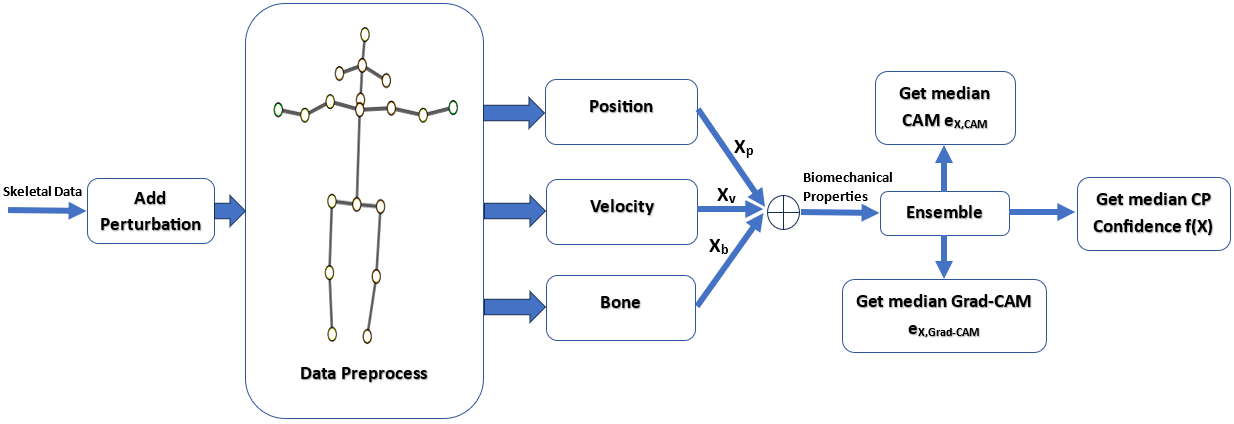}
        \caption{\label{fig:flow}}
    \end{subfigure}
    \vfill
    \begin{subfigure}[t]{0.8\textwidth}
        \centering
        \includegraphics[width=\textwidth]{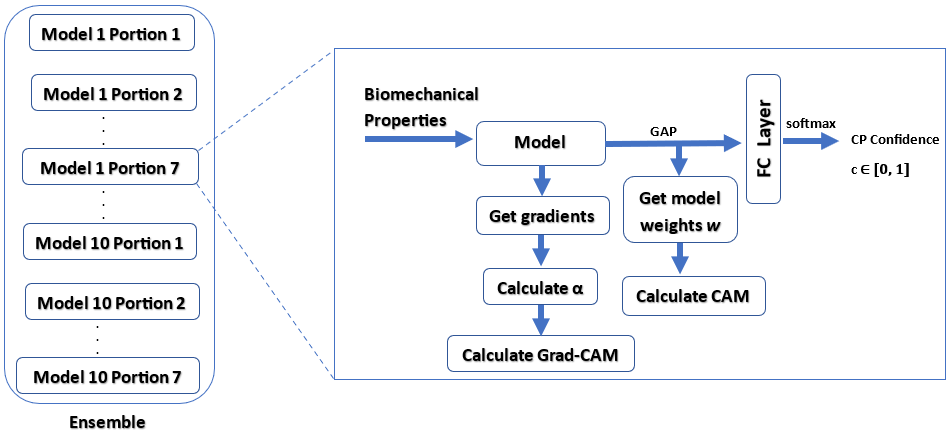}
        \caption{\label{fig:ensemble}}
    \end{subfigure}
    \caption{Overview of CP prediction ensemble pipeline, showing
    (\protect\subref{fig:flow})
    where the CP risk result and XAI attribution scores are obtained, and where the data perturbation is added for this study.
    The figure in (\protect\subref{fig:ensemble})
    shows the flow of data in each model in the CP prediction ensemble, and where the individual XAI attributions and CP risk result are obtained. The attribution scores are calculated from Equations \ref{eq:cam} and \ref{eq:gradcam}.}
    \label{fig:methods}
\end{figure*}

We evaluated the previously developed GCN-based CP prediction ensemble by Groos et al. \cite{groos2022development}, which has reported accuracy of 90.6\%, sensitivity of 71.4\%, and specificity of 94.1\% for classifying CP risk in infants based on videos. Before prediction, the raw skeletal data undergoes a preprocessing stage inspired by the methodology in \cite{song2022constructing}, which involves resampling to a consistent frame rate, applying motion smoothing, centering the coordinates around the pelvis, and normalizing bone lengths to account for variations in infant sizes. Next, the position, velocity, and bone length (distances between key body points) features for each joint are calculated. These features are organized into three parallel input branches (one for each), which the ensemble model processes separately. Fig.~\ref{fig:flow} shows this complete CP prediction pipeline and illustrates how these branches feed into the GCN architecture.

The GCN architecture used to analyze the three biomechanical movement properties was optimized through an automatic Neural Architecture Search (NAS) approach, which explored different architectural configurations. This yielded 10 distinct GCN architectures, each trained on 7 different subsets of the dataset, resulting in a total of 70 unique model instances, as shown in Fig.~\ref{fig:ensemble}. The final Ensemble-NAS-GCN model aggregates the predictions from these 70 GCN instances to enhance robustness and performance. Comprehensive details regarding the training process and dataset characteristics are available in the original study \cite{groos2022development}. Additional information about each of the 10 GCN model architectures can be found in Appendix Table \ref{tab:gcn_architectures}.

\subsection{Infant Skeletal Dataset}
\label{sec:dataset}
This study uses the same dataset from Groos et al. \cite{groos2022development}, comprising video recordings of 557 high-risk infants in Norway, Belgium, India, and the US. Recordings were collected during the FM period (9--18 weeks post-term age) following Prechtl's GMA protocols. CP diagnoses were confirmed by pediatricians at or after 12 months corrected age using the European classification system \cite{cans2000surveillance}. The dataset was split into 75\% training/validation and 25\% test sets, and we used the skeletal coordinates extracted via the pose estimation pipeline in \cite{groos2022towards}.

We further grouped the data into four risk categories based on the ensemble model's predictions for each 5-second window extracted from each infant's video. In this context, a window refers to a segment of skeletal data frames lasting 5 seconds, with each window overlapping the previous one by 2.5 seconds. The data grouping was determined by calculating the median risk score from the ensemble predictions for each window and computing the 25th and 75th percentiles to establish confidence intervals. Windows with an upper percentile below the predefined prediction threshold were labeled ``very low risk,” and those with a lower percentile above the threshold were labeled ``very high risk.” Windows with overlapping intervals (indicating some uncertainty) were classified ``low risk” or ``high risk” and excluded from this study to focus only on the most definitive cases for subsequent analysis. As a result of this data elimination process, we obtained 26,294 very low and 5,204 very high risk windows in the training/validation sets, and 5,782 very low risk and 987 very high risk windows in the test set. To manage testing time with our computational resources, we randomly selected about half of the available very high risk windows in the test set, resulting in 493 non-overlapping windows. We also drew an equal number of very low risk windows. Sampling was proportional to video length to preserve the natural distribution of data across varying durations.

\subsection{Explainable AI Methods}
\subsubsection{Class Activation Mapping (CAM)}

CAM was originally developed for convolutional neural networks (CNNs) by highlighting the pixels in an image that significantly influence the model's predictions \cite{zhou2016learning}. A heatmap layered over an image is obtained by multiplying the class-specific weights from the output layer with the feature maps from the final convolutional layer. Mathematically, this is represented as:

\begin{equation}
e_{\text{CAM}}(X) = \sum_n w_n^{\text{class}} F^n,
\label{eq:cam}
\end{equation}

where $e_{\text{CAM}}(X)$ is the activation map for input $X$, $w_n^{\text{class}}$ denotes the weights for the target class, and $F^n$ are the feature maps. Since Equation \ref{eq:cam} generalizes to any convolution-based architecture, CAM can be applied to GCNs as illustrated in \cite{song2022constructing} for a human activity recognition (HAR) model and used to identify influential joints or nodes in skeletal graphs.

\subsubsection{Gradient-weighted Class Activation Mapping (Grad-CAM)}

Grad-CAM extends the capabilities of CAM by overcoming its architectural constraints, particularly the need for a Global Average Pooling (GAP) layer and a fully connected classification layer \cite{selvaraju2020grad}. This is achieved by using the gradients of any target class flowing into any convolutional layer to generate a localization map that highlights important regions in the input data. The Grad-CAM activation map is computed as:

\begin{equation}
e_{\text{Grad-CAM}}(X) = \sum_n \alpha_n^{\text{class}} F^n,
\label{eq:gradcam}
\end{equation}

where \( \alpha_n^{\text{class}} \) represents the gradient-based importance weights for the \( n \)-th feature map \( F^n \). Like CAM, Grad-CAM can be applied to more than just image classifiers. For instance, Das et al. \cite{das2022gradient} demonstrated its effectiveness in spatio-temporal GCNs with skeletal data inputs.

\begin{figure*}[htbp]
  \centering
  \begin{tabular}{@{}c@{}c@{}}
    \begin{minipage}{.75\textwidth}
      \centering
      \includegraphics[scale=0.6]{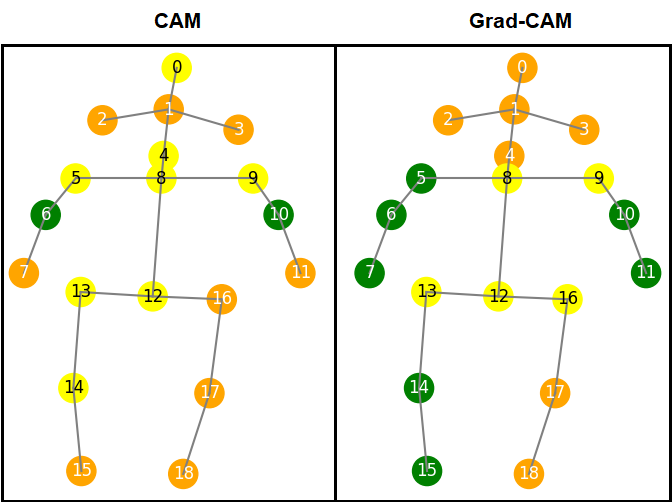}
    \end{minipage} & 
    \begin{minipage}{.25\textwidth}
      \centering
      \footnotesize
      \setlength{\tabcolsep}{2pt} 
      \renewcommand{\arraystretch}{1.2} 
      \begin{tabular}{@{}cccc@{}}
      \toprule
      k & CAM & Grad-CAM \\
      \midrule
      0 & 0.3241 & 0.3984 \\
      1 & 0.4782 & 0.3830 \\
      2 & 0.5126 & 0.5079 \\
      3 & 0.3617 & 0.3158 \\
      4 & 0.2985 & 0.4652 \\
      5 & 0.2794 & 0.0 \\
      6 & 0.0 & 0.0 \\
      7 & 0.4953 & 0.0 \\
      8 & 0.3212 & 0.0935 \\
      9 & 0.2847 & 0.1182 \\
      10 & 0.0 & 0.0 \\
      11 & 0.5291 & 0.0 \\
      12 & 0.3174 & 0.1013 \\
      13 & 0.2538 & 0.1278 \\
      14 & 0.3019 & 0.0 \\
      15 & 0.3746 & 0.0 \\
      16 & 0.4625 & 0.1424 \\
      17 & 0.3927 & 0.3360 \\
      18 & 0.5134 & 0.3067 \\
      \bottomrule
      \end{tabular}
    \end{minipage}
  \end{tabular}
  \caption{(\textbf{left}) Sample visualization of attribution scores from the ensemble model's XAI methods tested on the same video snippet. Green indicates very low attribution scores, yellow for low scores, orange for high scores, and red (not present in this example) for very high scores relative to the defined threshold scores (about 0.29 for CAM and 0.17 for Grad-CAM). (\textbf{right}) Numerical values of attribution scores, with \(k\) denoting the body keypoint number.}
  \label{fig:skeleton}
\end{figure*}

\subsubsection{Attribution Score Visualization}

Attribution scores range from 0 to 1 and are unitless, which make them inherently abstract. Lower scores imply more typical movement, whereas higher scores suggest atypical (possibly CP-related). However, these raw values are difficult to interpret without context. To enable clinician-friendly interpretation of CAM/Grad-CAM scores, we adopt a classification approach similar to the CP risk grouping in Section \ref{sec:dataset}. First, we aggregate the attribution scores from each model in the prediction ensemble for each joint, then compare their interquartile range (IQR) against a threshold \(\theta\), and color-code them. These thresholds were derived following the methodology of Groos et al. \cite{groos2022development}, where the CP risk threshold was first calibrated to achieve 70\% sensitivity matching GMA observers. The CAM and Grad-CAM thresholds were then set as the minimum mean attribution scores across keypoints for 70\% of infants classified with CP in the test set, yielding thresholds of approximately 0.29 for CAM and 0.17 for Grad-CAM.

The color coding of attribution score IQR with respect to \(\theta\) is as follows: 
\begin{itemize}
    \item \textbf{Green (``very low'')}: The entire IQR lies below \(\theta\), indicating minimal contribution to CP risk.
    \item \textbf{Yellow (``low'')}: The median is below \(\theta\), but part of the IQR is above \(\theta\), suggesting some uncertainty.
    \item \textbf{Orange (``high'')}: The median is above \(\theta\), but part of the IQR is below \(\theta\), again indicating uncertainty.
    \item \textbf{Red (``very high'')}: The entire IQR lies above \(\theta\), implying a strong contribution to CP risk.
\end{itemize}

Fig.~\ref{fig:skeleton} illustrates this visualization scheme alongside the corresponding raw numerical scores for comparison. This classification system will be revisited in Section~\ref{sec:top-k}, where it plays a crucial role in distinguishing significant and non-significant keypoints.

\section{Methods}

\begin{figure*}[htbp]
    \centering
    \begin{subfigure}[t]{0.8\textwidth}
        \centering
        \includegraphics[width=\textwidth]{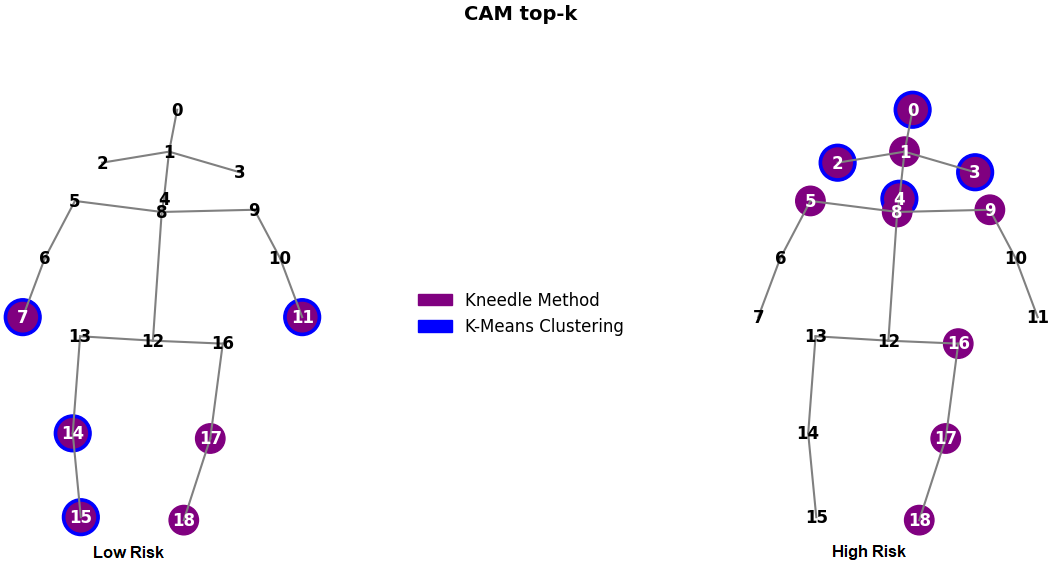}
        \caption{\label{fig:cam_top-k}}
    \end{subfigure}
    \vfill
    \begin{subfigure}[t]{0.8\textwidth}
        \centering
        \includegraphics[width=\textwidth]{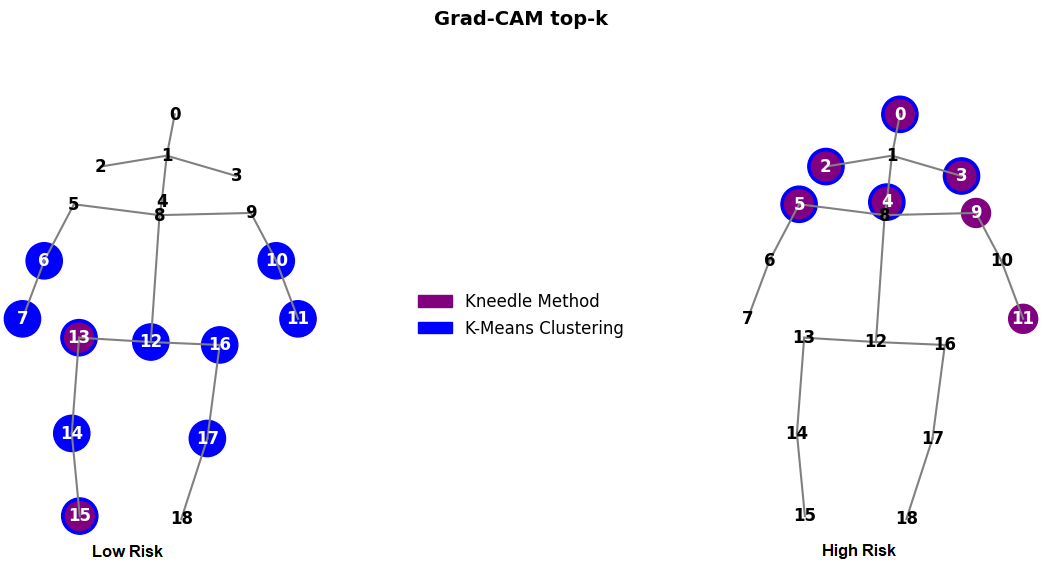}
        \caption{\label{fig:gcam_top-k}}
    \end{subfigure}
    \caption{The set of \emph{top-k} and \emph{non-top-k} joints identified using
    (\protect\subref{fig:cam_top-k})
    CAM,
    (\protect\subref{fig:gcam_top-k})
    Grad-CAM. Joints selected as significant by both the Kneedle Method and K-Means Clustering are classified as \emph{top-k}, while the remaining joints are categorized as \emph{non-top-k}. Although CAM and Grad-CAM differ in the specific joints they flag as significant, a consistent pattern emerges: in low-risk windows, the extremities (arms and legs) seem to be most influential, whereas in high-risk windows, the head, neck, and shoulders seem to carry more significance. These sets of \emph{top-k} and \emph{non-top-k} joints are used throughout the experiment.} 

     \label{fig:top-k-set}
\end{figure*}

Following the XAI metrics methodology in \cite{agarwal2022openxai}, we first identify \emph{top-k} and \emph{non-top-k} features in the input data. \emph{Top-k} refers to the joints contributing significantly to CP risk classification, while \emph{non-top-k} are those with minimal impact. In very low risk infants, keypoints classified as green or yellow contribute to the prediction, whereas in very high risk infants, keypoints classified as orange or red are considered significant.
 
\subsection{Identifying Significant and Non-significant Joints}
\label{sec:top-k}

We focus only on 5-second windows classified as ``very low risk” or ``very high risk” as described in Section \ref{sec:dataset}. To identify \emph{top-k} and \emph{non-top-k} joints, we calculate the frequency of each joint receiving a green score in the low risk group, or a red score in the high risk group across all windows to rank joints by importance. The resulting feature importance percentages are shown in Fig.~\ref{fig:cam_gcam_lowrisk_percentages} and \ref{fig:cam_gcam_hirisk_percentages} in the Appendix. In the methodology proposed in \cite{agarwal2022openxai}, $k$ in \emph{top-k} is evaluated from 1 to the total number of features. However, predefining a single value of $k$ is necessary to perform our experiments. Although domain knowledge could guide in selecting significant features as illustrated in \cite{srikanth2022xai, palaniyappan2022aqx}, this approach is not as straightforward here, as clinical observation like the GMA focuses on holistic, gestalt evaluations of movement patterns \cite{aizawa2021general} rather than analyzing the contribution of individual body points.

We thus applied two methods to determine $k$ objectively. First is Knee Point Detection, which plots feature importance in descending order and uses the Kneedle algorithm \cite{satopaa2011finding, Arvai_kneed_2020} to identify the “knee,” where additional features add diminishing returns. Second is clustering, which performs k-means with two clusters to split features into “significant” and “non-significant.” The cluster with the higher centroid contains the \emph{top-k}. Next, we implemented a conservative voting system: a joint is classified as \emph{top-k} only if both methods agree. The remaining joints are then categorized as \emph{non-top-k}. The algorithm for this selection process is detailed in Appendix Section \ref{alg:feature_significance}, and the final \emph{top-k} and \emph{non-top-k} groupings from the training/validation set are shown in Fig.~ \ref{fig:top-k-set}. This is used in subsequent perturbation tests on the test set.

\subsection{Velocity Perturbation}
\begin{figure}[htbp]
    \centering
    \includegraphics[scale=0.3]{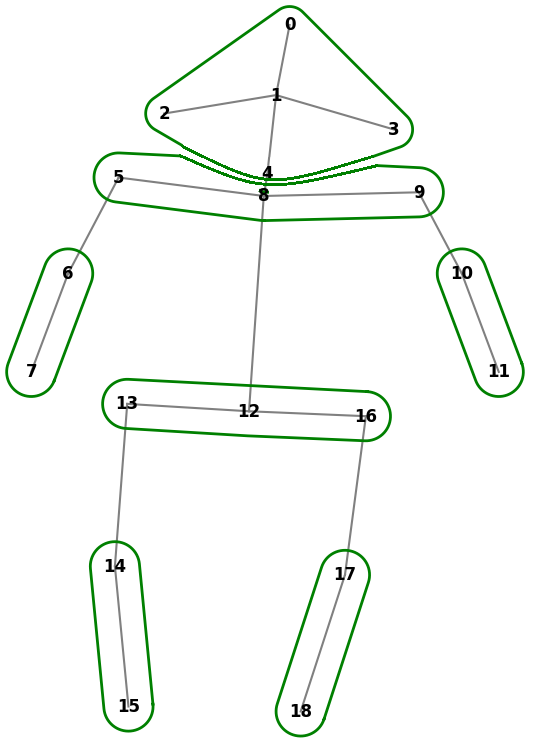}
    \caption{Definition of infant skeletal segments applied in velocity perturbation. If one keypoint in a segment is selected for perturbation, the entire segment is perturbed together to preserve bone length and maintain anatomically realistic movement.}
    \label{fig:segments}
\end{figure}

The velocity input feature to the prediction model is defined as the difference in joint positions between consecutive frames. Since the sampling rate is 30 frames per second, velocity is measured in units of normalized position changes per 1/30 second. Perturbing the velocity involves systematically increasing or decreasing this rate of change in position to analyze their impact on model predictions. The perturbation is introduced at the data preprocessing stage, as illustrated in Fig.~\ref{fig:flow}.

The perturbation process consists of four stages:

\begin{enumerate}
    \item \textbf{Reference Distribution Calculation}: For each risk group in the training/validation set, compute the 5th (P5\_reference) and 95th (P95\_reference) percentile velocities per joint and use them as typical range.
    
    \item \textbf{Sample-Specific Velocity Scaling}: For each test window, calculate sample-specific velocity ranges in each joint (P5\_sample and P95\_sample), then compute:
    \begin{enumerate}
        \item minimum scaling factor as \( \frac{\text{P5\_reference}}{\text{P5\_sample}} \), and
        \item maximum scaling factor as \( \frac{\text{P95\_reference}}{\text{P95\_sample}} \).
    \end{enumerate}
    
    \item \textbf{Biomechanically-Constrained Perturbation}: Joints are grouped into ``segments” (Fig.~\ref{fig:segments}) to preserve realistic limb length and motion patterns. In slowdown mode (P5-based scaling), apply the largest factor \(\le 1\); in speedup mode (P95-based scaling), apply the smallest factor \(\ge 1\).
    
    \item \textbf{Velocity Modification and Evaluation}: For both \emph{top-k} and \emph{non-top-k} joints, we apply slowdown multipliers 
    \(\{0.20, 0.25, 0.33, 0.5, 1\}\times\text{(min factor)}\) 
    and speedup multipliers 
    \(\{1, 2, 3, 4, 5\}\times\text{(max factor)}\).
    Each perturbed sample is fed into the model to record changes in CP risk scores.
\end{enumerate}

The velocity modification is applied using temporal interpolation of joint trajectories. This is implemented in code using the interp1d function from SciPy library \cite{jones2001scipy}. We performed eight perturbation analyses, each combining an XAI method (CAM or Grad-CAM), a risk group (low or high), and a feature subset (\emph{top-k} or \emph{non-top-k}). The algorithmic summary of this process is shown in Appendix \ref{alg:velocity_perturbation}.

\subsection{Angular Perturbation}
\label{sec:angular_perturbation}
An angle $\Delta$ is defined as the frame-by-frame change in a joint’s angle relative to the horizontal axis when connected to its adjacent ``parent" joint. Angle perturbation is done by multiplying \( \Delta \) by a scaling factor while maintaining the original bone length. Although modifying $\Delta$ indirectly affects the velocity and position of joints, angle perturbation differs fundamentally by expanding or contracting the repertoire of joint positions while still preserving anatomical plausibility. In contrast, velocity perturbation preserves the original set of positions but only alters the timing of transitions between them.

The perturbation process follows similar stages as velocity perturbation, but retaining only the head segment (joints 0 to 4) as a single rigid unit since other joints have more freedom to move independently with respect to their parent joint. The algorithmic summary of this process is shown in Appendix \ref{alg:angle_perturbation}, including the geometric transformation process to maintain frame-by-frame bone lengths. We conduct eight distinct angular perturbation analyses, yielding the same set of results as the velocity perturbation experiment.

\subsection{Combined Perturbation}

We apply a sequential perturbation of velocities and angles of \emph{top-k} and \emph{non-top-k} joints. The effect is that velocity slowdown is followed by angular reduction, or velocity speedup is followed by angular amplification. This combined perturbation process extends the individual analyses to provide additional insights into the interplay of velocity and angular changes in the context of CP risk prediction.

\begin{figure*}[htp]
    \centering
    \begin{subfigure}[t]{\textwidth}%
        \centering
        \includegraphics[width=\textwidth, height=0.4\textheight, keepaspectratio]{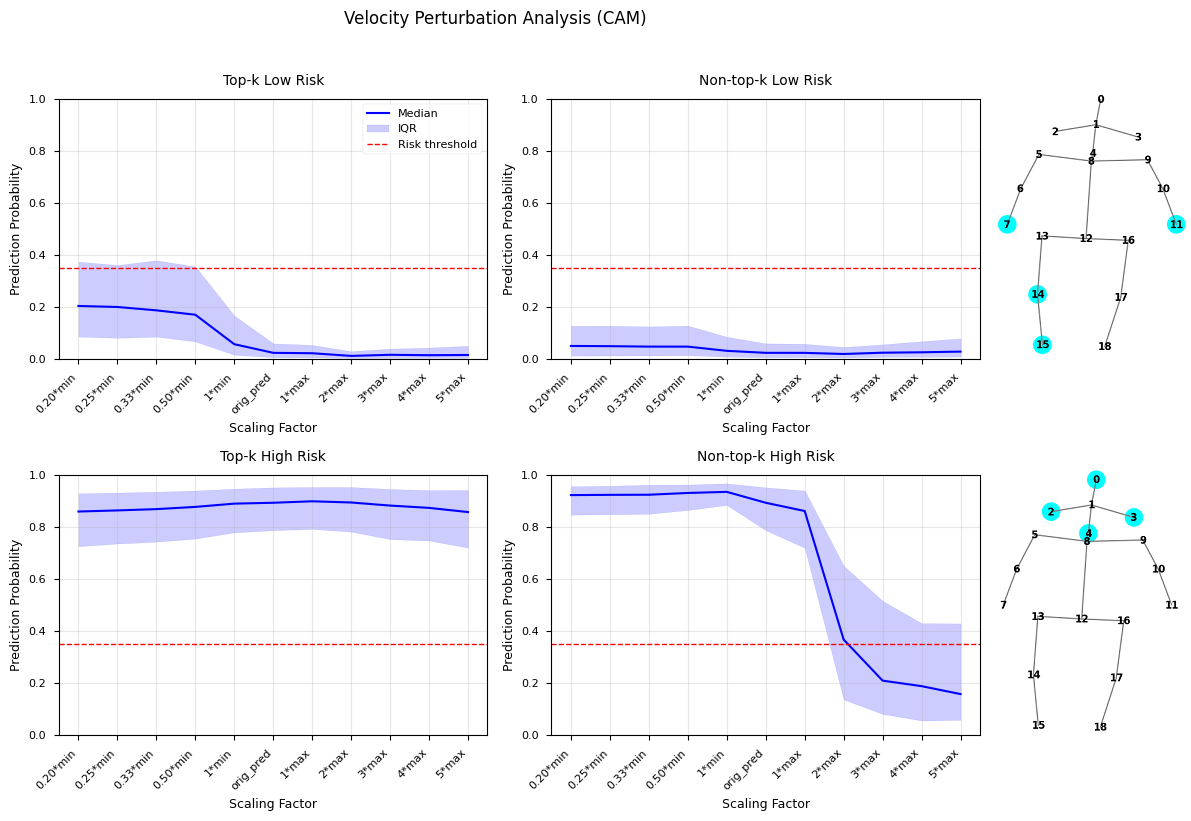}
        \caption{\label{fig:cam_velocity}}
    \end{subfigure}

    \vspace{0.5cm}  

    \begin{subfigure}[t]{\textwidth}%
        \centering
        \includegraphics[width=\textwidth, height=0.4\textheight, keepaspectratio]{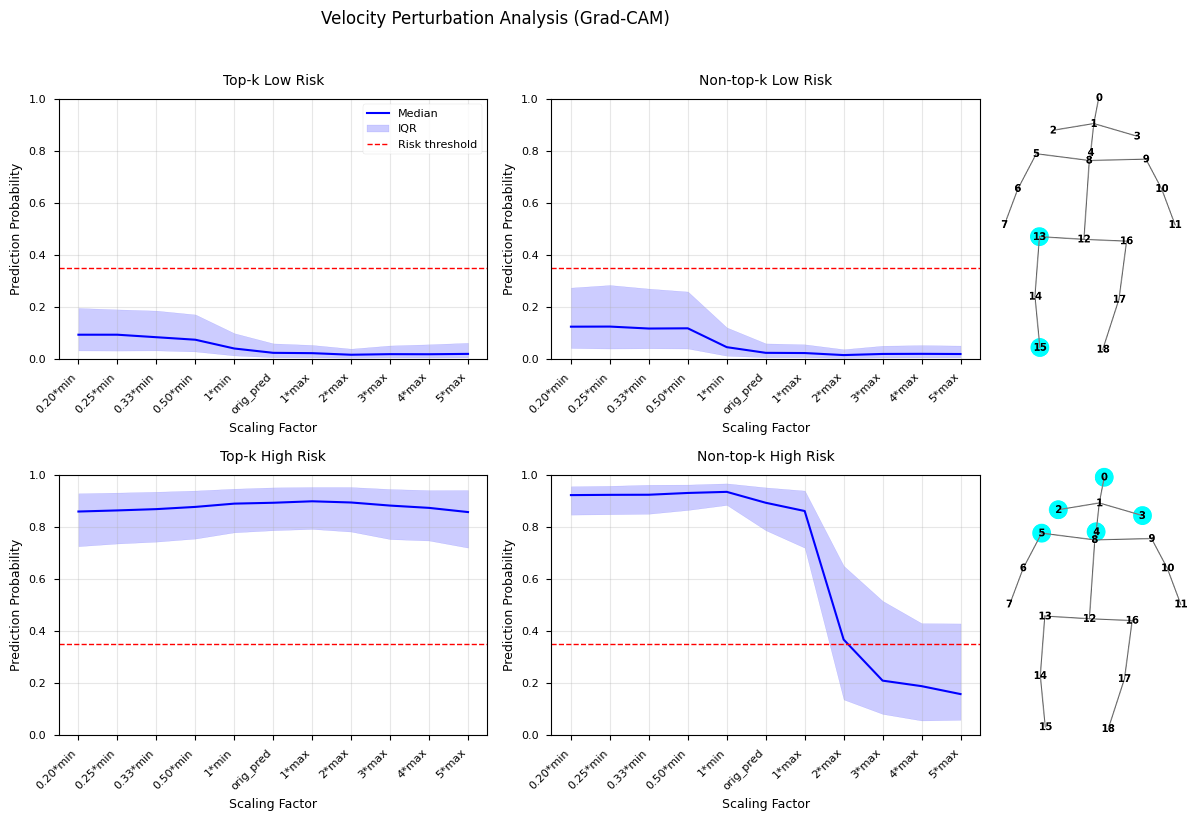}
        \caption{\label{fig:gcam_velocity}}
    \end{subfigure}
    \caption{Velocity perturbation analysis under \protect\subref{fig:cam_velocity}) CAM and \protect\subref{fig:gcam_velocity}) Grad-CAM. Each sub-plot shows how scaling the velocity of either \emph{top-k} (highlighted in teal in the skeletal diagram on the right) or \emph{non-top-k} joints affects the model’s predicted CP risk for low and high risk windows, with the median prediction (solid line) and interquartile range (blue shaded region). The horizontal dashed line represents the model’s CP risk threshold. $min$ refers to the 5th-percentile velocity, while $max$ refers to the 95th-percentile velocity.
    }
    \label{fig:velocity_comparison}
\end{figure*}

\begin{figure*}[htp]
    \centering
    \begin{subfigure}[t]{\textwidth}%
        \centering
        \includegraphics[width=\textwidth, height=0.4\textheight, keepaspectratio]{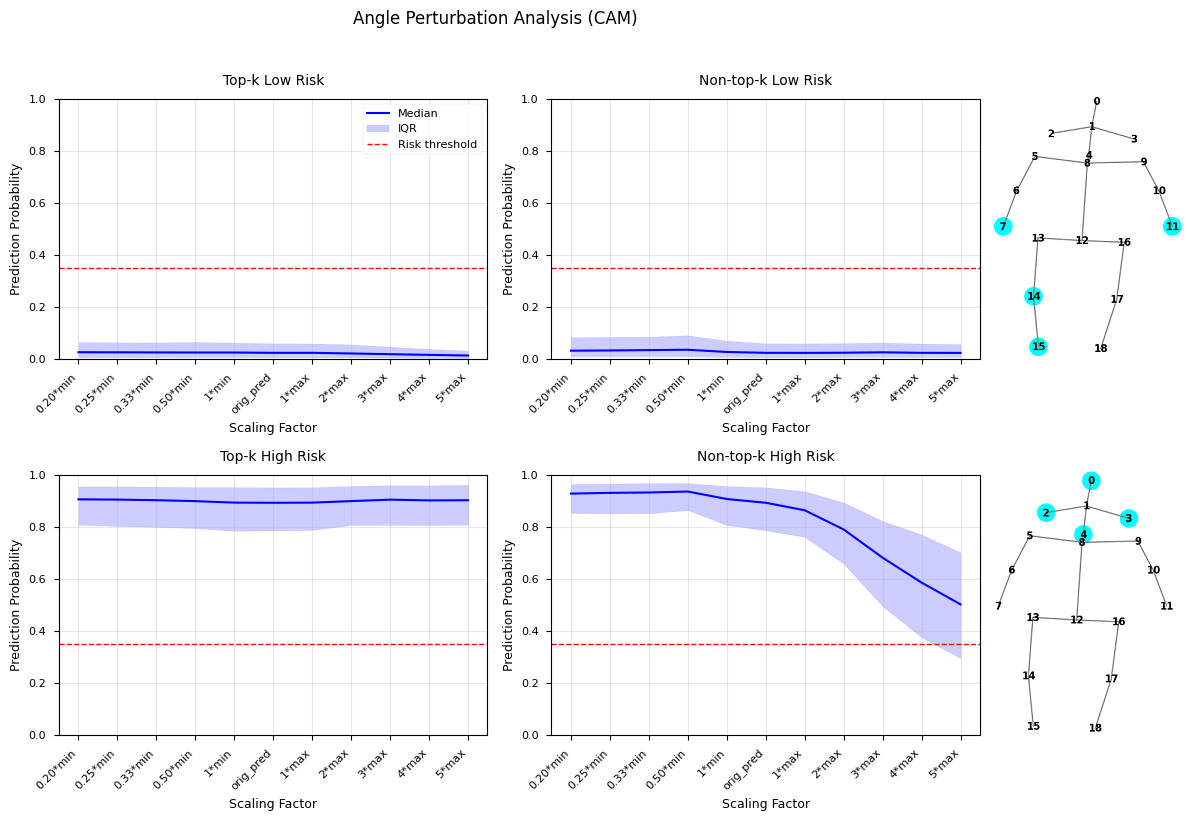}
        \caption{\label{fig:cam_angle}}
    \end{subfigure}

    \vspace{0.5cm}  

    \begin{subfigure}[t]{\textwidth}%
        \centering
        \includegraphics[width=\textwidth, height=0.4\textheight, keepaspectratio]{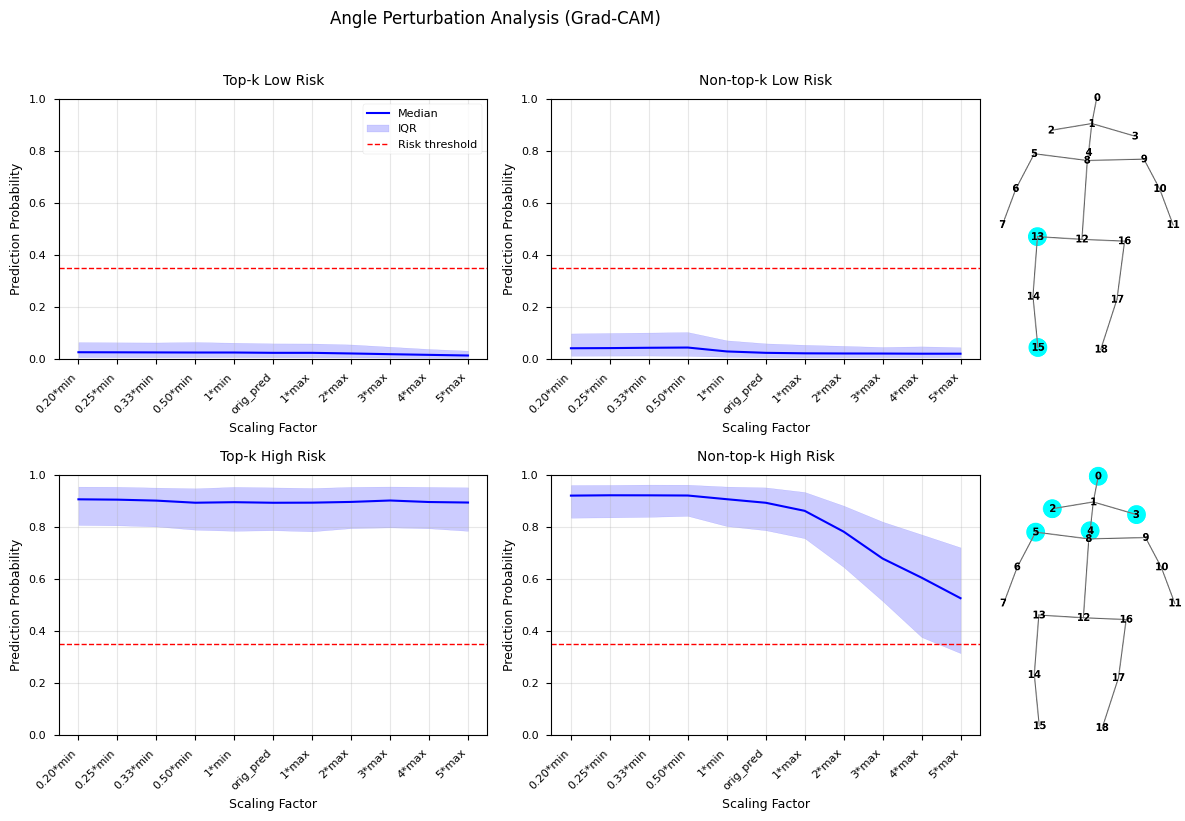}
        \caption{\label{fig:gcam_angle}}
    \end{subfigure}
    \caption{Angle perturbation analysis under \protect\subref{fig:cam_angle}) CAM and \protect\subref{fig:gcam_angle}) Grad-CAM. Each sub-plot shows how scaling the angular changes of either \emph{top-k} (highlighted in teal in the skeletal diagram on the right) or \emph{non-top-k} joints affects the model’s predicted CP risk for low and high risk windows, with the median prediction (solid line) and interquartile range (blue shaded region). The horizontal dashed line represents the model’s CP risk threshold. $min$ refers to the 5th-percentile $\Delta$, while $max$ refers to the 95th-percentile $\Delta$.
    }
    \label{fig:angle_comparison}
\end{figure*}

\begin{figure*}[htp]
    \centering
    \begin{subfigure}[t]{\textwidth}%
        \centering
        \includegraphics[width=\textwidth, height=0.4\textheight, keepaspectratio]{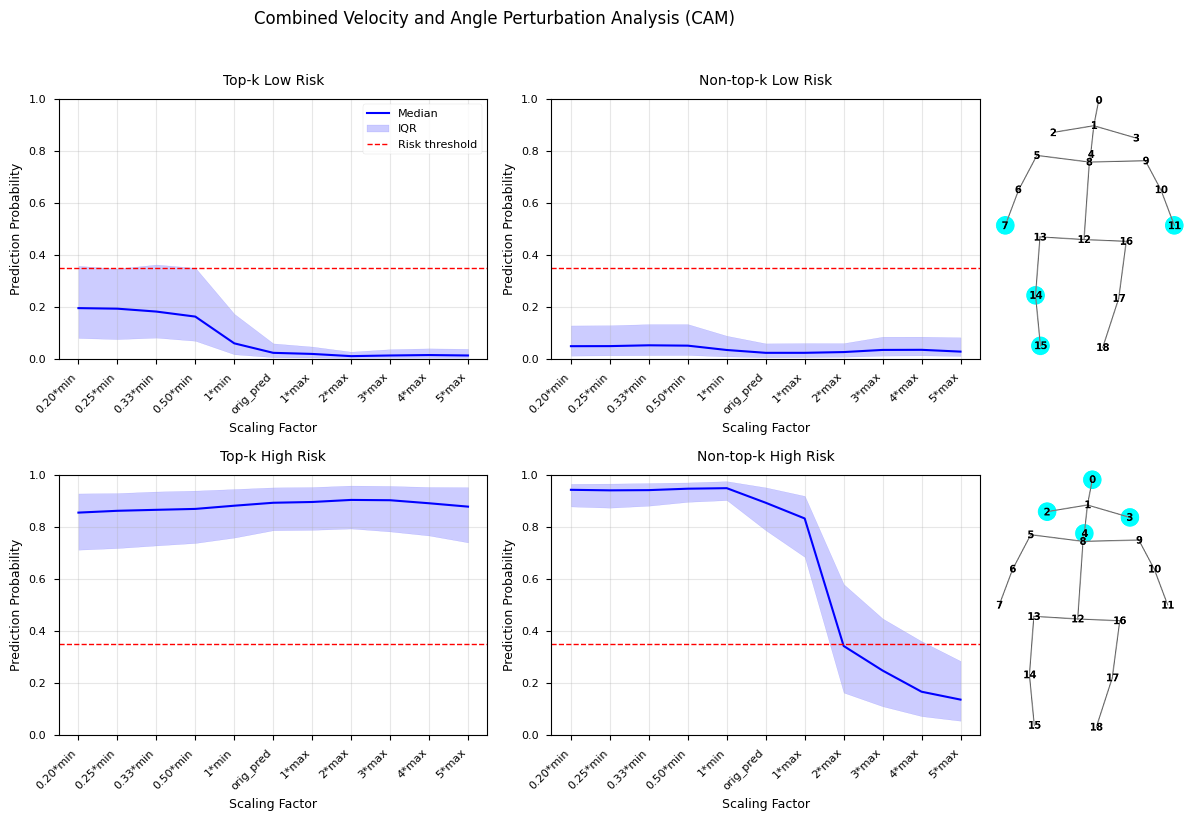}
        \caption{\label{fig:cam_combined}} 
    \end{subfigure}

    \vspace{0.5cm}  

    \begin{subfigure}[t]{\textwidth}%
        \centering
        \includegraphics[width=\textwidth, height=0.4\textheight, keepaspectratio]{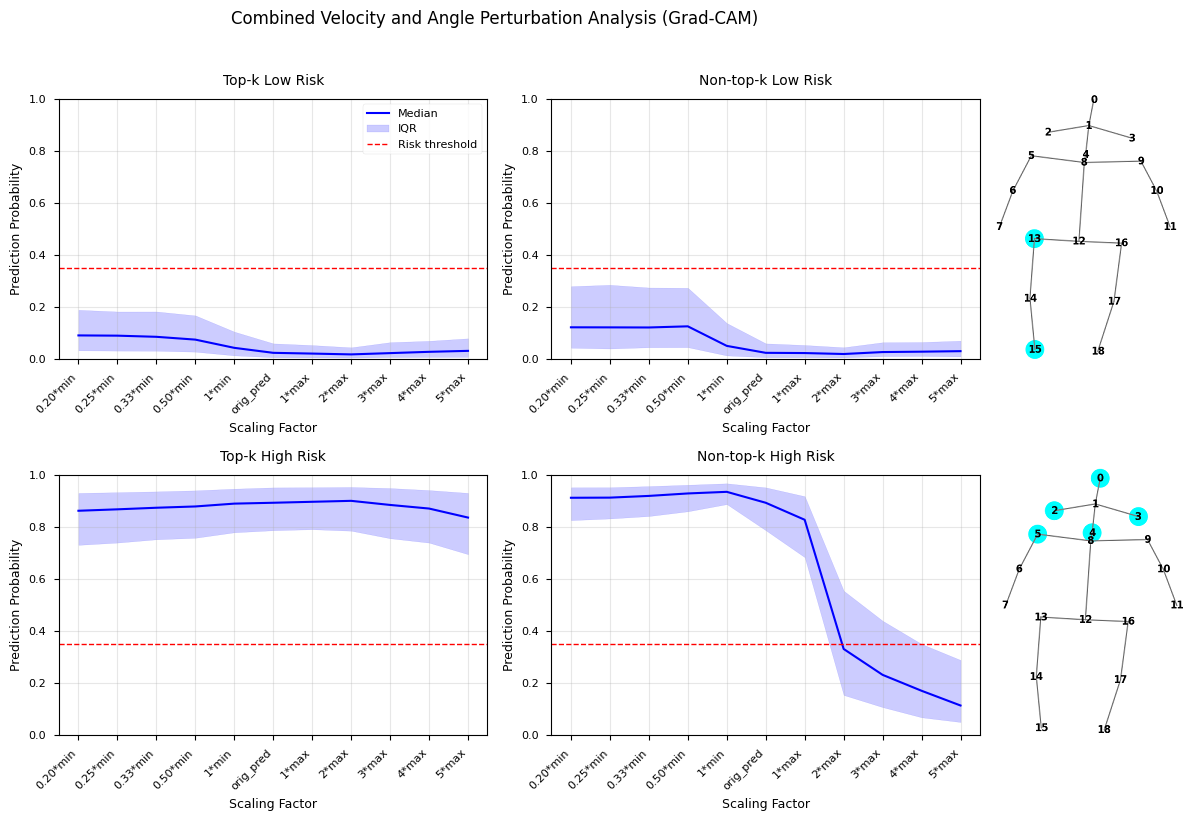}
        \caption{\label{fig:gcam_combined}}
    \end{subfigure}
    \caption{Combined velocity + angle perturbation analysis under \protect\subref{fig:cam_combined}) CAM and \protect\subref{fig:gcam_combined}) Grad-CAM. Each sub-plot shows how the subsequent scaling of velocity and angular changes of either \emph{top-k} (highlighted in teal in the skeletal diagram on the right) or \emph{non-top-k} joints affects the model’s predicted CP risk for low and high risk windows, with the median prediction (solid line) and interquartile range (blue shaded region). The horizontal dashed line represents the model’s CP risk threshold. $min$ refers to the 5th-percentile velocity and $\Delta$, while $max$ refers to the 95th-percentile velocity and $\Delta$.
    }
    \label{fig:combined_comparison}
\end{figure*}

\section{Results and Discussion}

\subsection{Velocity Perturbation}
Fig.~\ref{fig:velocity_comparison} provides an overview of how velocity perturbations affect CP risk predictions under CAM (\ref{fig:cam_velocity}) and Grad-CAM (\ref{fig:gcam_velocity}). Below, we discuss these patterns in detail.

\subsubsection{CAM and Grad-CAM}

For CAM analysis in low risk windows, slowing the \emph{top-k} joints (i.e., arms and right leg) raises the median risk from $\mathord{\sim}0.02$ to 0.20, suggesting the model associates slower movements with higher CP risk. Conversely, speeding them up lowers the risk to $\mathord{\sim}0.01$–0.02, reinforcing a low risk assessment. Perturbing \emph{non-top-k} joints (i.e., all except arms and right leg) has a smaller effect, raising risk from 0.02 to $\mathord{\sim}0.05$ when slowed. In high risk windows, altering the \emph{top-k} joints (head segment) has minimal impact, while modifying \emph{non-top-k} joints produces very noticeable effects. Slowing them slightly increases risk (0.89 to $>$0.92), whereas speeding them up reduces risk significantly, dropping it to 0.36–0.15.

Grad-CAM follows a broadly similar pattern. In low risk windows, slowing the \emph{top-k} joints (i.e., right leg and right hip) raises risk to $\mathord{\sim}0.09$, while speeding them lowers it to $\mathord{\sim}0.02$. Perturbing \emph{non-top-k} joints (i.e., all except the right leg and right hip) also increases risk if slowed ($\mathord{\sim}0.12$) and reduces it if sped up ($\mathord{\sim}0.02$). In high risk windows, modifying \emph{top-k} joints (i.e., head, neck, shoulders) has little effect. However, \emph{non-top-k} joints (i.e., all except the head, neck, and shoulders) exert a very strong influence: slowing them raises risk to $\mathord{\sim}0.93$, while speeding them lowers it to below $\mathord{\sim}0.20$.

\subsubsection{Potential Clinical Interpretations of Velocity Perturbations}

A consistent pattern emerges regardless of whether CAM or Grad-CAM is used: slower movements in the arms, legs, and hips tend to elevate the model’s CP risk estimate, whereas faster movements drive the risk down. In high risk windows, increasing the velocity of these joints substantially lowers the model’s predicted CP risk. In contrast, slowing down the limbs in low risk windows raises the risk somewhat but does not completely shift the prediction to high risk, implying that there are other factors beyond velocity indicative of atypical movement. Future research could explore whether these technical findings relate to some qualitative descriptions of these joints, such as variable acceleration in FMs \cite{einspieler2019cerebral, einspieler2016fidgety}, monotonous kicking \cite{bruggink2009early}, and cramped-synchronized movement \cite{bruggink2009early, yuge2011movements, yang2012cerebral}.

\subsection{Angular Perturbation}
Fig.~\ref{fig:angle_comparison} provides an overview of how angular perturbations affect CP risk predictions under CAM (\ref{fig:cam_angle}) and Grad-CAM (\ref{fig:gcam_angle}). Below, we discuss these patterns in detail.

\subsubsection{CAM and Grad-CAM}

For CAM analysis in low risk windows, altering angles in \emph{top-k} joints causes only minor risk fluctuations (e.g., from 0.02 to 0.03 when angles are drastically reduced), while further increases slightly lower the risk. In \emph{non-top-k} joints, the effect is slightly larger but remains moderate (0.02 to $\mathord{\sim}0.03 - 0.04$). Thus, the model is less sensitive to angular changes than velocity modifications in low risk cases. In high risk windows, \emph{top-k} angular perturbations barely shift the median risk from $\mathord{\sim}0.89 - 0.90$. However, altering angles in \emph{non-top-k} joints can raise risk above $\mathord{\sim}0.92$ if reduced or lower it toward $\mathord{\sim}0.50$ if increased.

Grad-CAM shows a similar trend. In low risk windows, \emph{top-k} joints are relatively robust to angular scaling, with changes within 0.01–0.02 of the baseline. \emph{Non-top-k} joints exhibit a slightly noticeable response, though still less significant than with velocity perturbations. In high risk conditions, altering \emph{top-k} joints results in minimal shifts in the risk score, while substantial changes in \emph{non-top-k} angles lead to much notable risk increases or decreases (e.g., from $\mathord{\sim}0.89$ up to 0.92, or down to 0.68–0.52).

\subsubsection{Potential Clinical Interpretations of Angular Perturbations}

As mentioned in Section \ref{sec:angular_perturbation}, angular perturbation involves expanding or contracting an infant’s range of possible joint positions. In general, this approach produces smaller changes in the model’s CP risk estimates compared to velocity modifications, except when angles in \emph{non-top-k} joints (e.g., arms, legs, and hips) are increased in high risk windows. These findings suggest that to some extent, a larger range of these joints' positions may be viewed by the model as indicative of more typical movement. It remains to be explored whether this technical observation could be associated with GMA-based qualitative descriptions such as the increased movement variability \cite{einspieler2016fidgety, einspieler2008human} that typically accompanies FMs.

\subsection{Combined Velocity and Angular Perturbation}
Fig.~\ref{fig:combined_comparison} provides an overview of how the combined velocity and angular perturbations affect CP risk predictions under CAM (\ref{fig:cam_combined}) and Grad-CAM (\ref{fig:gcam_combined}). Below, we discuss these patterns in detail.

\subsubsection{CAM and Grad-CAM}

When velocity and angle are perturbed sequentially, the effects of individual perturbations are slightly amplified. For CAM analysis in low risk windows, slowing and contracting \emph{top-k} joints raises risk more substantially (e.g., from 0.02 to $\mathord{\sim}0.20$), while speeding and expanding them further reinforces low risk predictions. \emph{Non-top-k} perturbations also affect risk but less dramatically. In high risk windows, modifying \emph{top-k} joints results in minor shifts around $\mathord{\sim}0.89$, whereas large \emph{non-top-k} changes can push risk to $\mathord{\sim}0.95$ or drop it to 0.14–0.17 with extreme velocity-angle increases.

Grad-CAM under combined perturbations similarly amplifies effects. In low risk windows, slowing and contracting \emph{top-k} joints raises risk to $\mathord{\sim}0.09$ (from 0.02), while speeding and expanding lowers it slightly ($\mathord{\sim}0.02$). \emph{Non-top-k} perturbations can elevate risk further (up to $\mathord{\sim}0.12$) but remain less extreme than high risk cases. In high risk windows, \emph{non-top-k} joints show notable sensitivity: decreased scaling can push risk above 0.91, while the opposite drastically reduces it, with some windows dropping below 0.20–0.30. Meanwhile, perturbing \emph{top-k} joints does not cause the same dramatic shifts in prediction.

\subsubsection{Potential Clinical Interpretations of Combined Perturbations}

Together, the combined perturbation experiments reinforce the findings that the model has high sensitivity to specific joint velocities, and a slight sensitivity to range of joint positions, especially in arms, hips, and legs. The magnitude of these effects is generally much larger in high risk windows. This suggests that faster and broader movements of these joints could be associated with typical development. 

\section{Conclusion, Limitations, and Future Directions}

Our findings highlight certain quantifiable biomarker-like features, particularly velocity- and position-related parameters in the arms, hips, and legs. However, prospective clinical validation studies are needed to confirm their practical utility. While these perturbation experiments offer initial insights into the model’s inner workings, they do not fully capture the biomechanical complexity of real infant movement. Future work could explore GMA-informed perturbation strategies to better approximate genuine infant motion. In particular, the Motor Optimality Score (MOS) which includes qualitative descriptions of arm, hip, and leg movements (e.g., swipes, kicking) \cite{einspieler2019cerebral, bertoncelli2024motor}, as well as other joints, may offer valuable insights. Additionally, perturbations to the head did not substantially alter the predicted risk, suggesting that our methods may not fully capture the complexity of head movements or that the model is inherently biased toward limb dynamics. Future investigations could refine head perturbations, as both CAM and Grad-CAM highlight their importance in high risk windows, yet the specific movement features driving this significance remain unclear from our experiments.

From the standpoint of explanation detail, both CAM and Grad-CAM show partial convergence in their attributions. CAM sometimes concentrates on a single side (e.g., the right leg), while Grad-CAM shows more asymmetry in areas such as the ankle and hip. Both nonetheless consistently identify the leg in low risk windows and the head in high risk windows. Similar patterns of narrower versus broader activation maps have been observed in image classification \cite{pinciroli2021comparing}. However, while image-based studies have ground truth, our study does not have such labels. Despite this limitation, the partial overlap suggests both methods capture important kinematic cues that distinguish typical from atypical movement. Meanwhile, their differences leave open the question of which method is better for skeleton-based human movement analysis. Overall, our results demonstrate that XAI-driven perturbation analysis effectively reveals how the CP prediction model makes decisions and holds promise for identifying disease-specific movement biomarkers, ultimately facilitating adoption of AI-assisted early CP prediction.

\bibliographystyle{IEEEtran}
\bibliography{ref}

\onecolumn
\clearpage
\appendix
\clearpage
\section{GCN Architectures}

\begin{table*}[htbp]
\centering
\caption{The architectures of the 10 models in the ensemble. Abbreviations: br., branch; bottl., bottleneck; MBC., mobile inverted bottleneck convolution; den., dense; mod., module; tmp., temporal; lin., linear; gl., global; sp., spatial; DA, disentangled aggregation; SC, spatial configuration; SE, Squeeze-and-Excitation; abs., absolute; ch., channel; ReLU, rectified linear unit; sw., swish; AUC, area under the receiver operating characteristic curve.}
\small
\begin{adjustbox}{max width=\textwidth}
\begin{tabular}{>{\arraybackslash}m{4cm}|>{\centering\arraybackslash}m{0.6cm}|>{\arraybackslash}m{0.6cm}|>{\centering\arraybackslash}m{0.6cm}|>{\centering\arraybackslash}m{0.6cm}|>{\centering\arraybackslash}m{0.6cm}|>{\centering\arraybackslash}m{0.6cm}|>{\centering\arraybackslash}m{0.6cm}|>{\centering\arraybackslash}m{0.6cm}|>{\centering\arraybackslash}m{0.6cm}|>{\centering\arraybackslash}m{0.6cm}}
\toprule
\textbf{Architectural choice} & 1 & 2 & 3 & 4 & 5 & 6 & 7 & 8 & 9 & 10 \\
\midrule
No. modules of input br. & 3 & 3 & 2 & 2 & 2 & 2 & 3 & 3 & 2 & 2 \\
Width of input br. & 10 & 10 & 12 & 10 & 8 & 6 & 8 & 6 & 12 & 12 \\
Block type in initial mod. & Bottl. & Basic & Basic & Basic & Bottl. & Basic & Basic & MBC. & Bottl. & Basic \\
Residual type in initial mod. & None & Den. & None & Block & Den. & Den. & Mod. & Block & Den. & Den. \\
No. tmp. scales in input br. & 1 & 3 & 2 & 2 & 3 & 2 & 2 & 1 & 2 & 2 \\
No. levels of main br. & 3 & 1 & 3 & 2 & 2 & 2 & 2 & 2 & 2 & 1 \\
No. modules of main br. levels & 3 & 2 & 1 & 1 & 3 & 3 & 2 & 3 & 1 & 1 \\
Width of first level of main br. & 12 & 12 & 12 & 10 & 12 & 12 & 10 & 12 & 8 & 12 \\
No. tmp. scales in main br. & 1 & 2 & 2 & 3 & 2 & 1 & 2 & 1 & 1 & 1 \\
Pooling layer type & Gl. & Gl. & Gl. & Sp. & Sp. & Gl. & Gl. & Gl. & Gl. & Sp. \\
Graph convolution type & DA 2 & DA 4+2 & SC & DA 4 & SC & DA 4 & DA 2 & DA 2 & DA 4 & SC \\
Block type & Basic & MBC. & Basic & Basic & Bottl. & Bottl. & Basic & Basic & Bottl. & Basic \\
Bottl. factor & 4 & 2 & 2 & 2 & 2 & 2 & 4 & 4 & 4 & 4 \\
Residual type & None & Block & Den. & None & None & Block & None & Den. & Block & None \\
SE type & None & Outer & Inner & None & Outer & None & None & Outer & Outer & None \\
SE ratio & - & 2 & 2 & 4 & 2 & 4 & 4 & 4 & 4 & 4 \\
SE ratio type & - & Abs. & Abs. & Abs. & - & Abs. & - & Abs. & Abs. & - \\
Attention type & Ch. & - & - & - & Ch. & - & Ch. & - & Ch. & - \\
Nonlinearity type & ReLU & Sw. &  ReLU & Sw. & Sw. & ReLU & ReLU & Sw. & ReLU & Sw. \\
Tmp. kernel size & 9 & 7 & 9 & 7 & 9 & 7 & 9 & 7 & 9 & 7 \\
AUC & 0.949 & 0.942 & 0.938 & 0.943 & 0.937 & 0.956 & 0.953 & 0.953 & 0.932 & 0.947 \\
\bottomrule
\end{tabular}
\label{tab:gcn_architectures}
\end{adjustbox}
\end{table*}

\clearpage
\section{Feature Importance Percentages}

\begin{minipage}{\textwidth}
    \centering
    \includegraphics[width=0.8\textwidth]{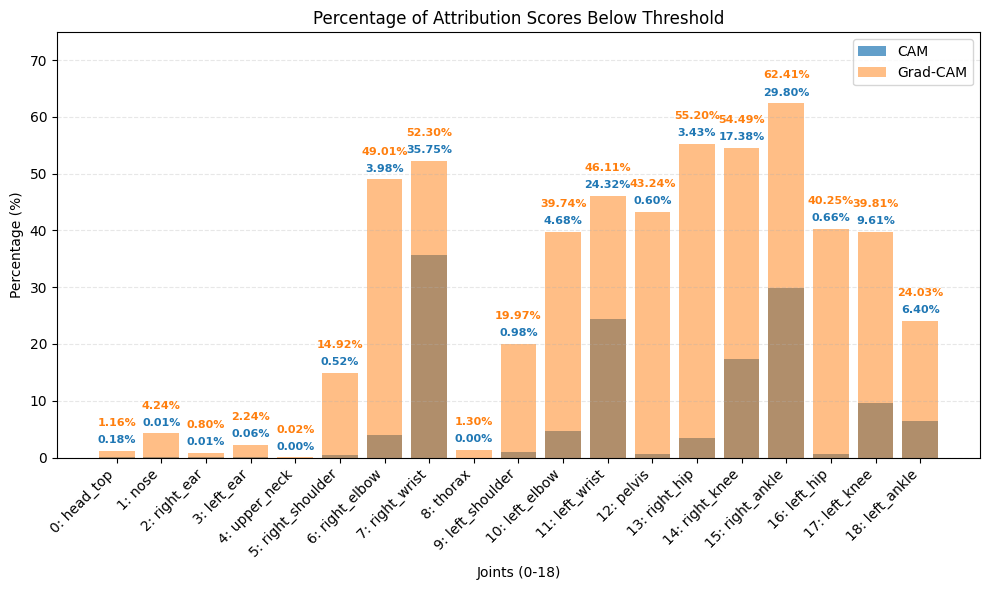}
    
    \captionof{figure}{Frequency of Green Joints from CAM and Grad-CAM on Low Risk Windows}
    \label{fig:cam_gcam_lowrisk_percentages}
\end{minipage}

\vspace{1cm} 

\begin{minipage}{\textwidth}
    \centering
    \includegraphics[width=0.8\textwidth]{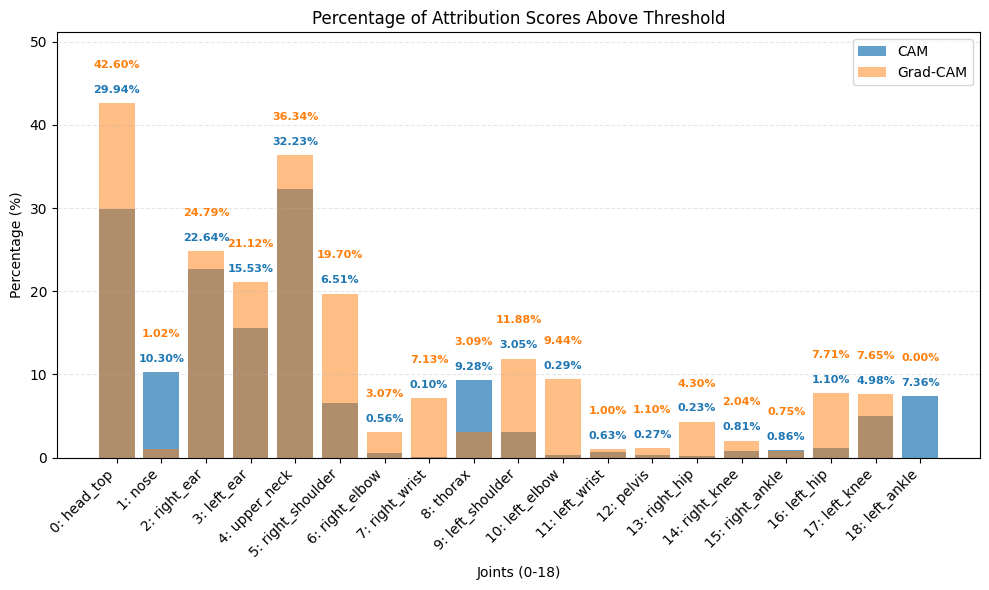}
    
    \captionof{figure}{Frequency of Red Joints from CAM and Grad-CAM on High Risk Windows}
    \label{fig:cam_gcam_hirisk_percentages}
\end{minipage}

\clearpage
\section{Algorithm}

\subsection{Feature Significance}
\label{alg:feature_significance}
\begin{algorithm*}[!ht]
\caption{Identification of $top$-$k$, $non$-$top$-$k$ Features}
\KwIn{
Training/validation dataset $D$ containing overlapping 5-second skeletal windows \
Attribution score thresholds $t_{CAM}$ = 0.29, $t_{GradCAM}$ = 0.17
}
\KwOut{
Sets of $top$-$k$ and $non$-$top$-$k$ joints for each XAI method and risk group: \\
1. $top\text{-}k_{CAM,low}$, $non\text{-}top\text{-}k_{CAM,low}$ \\
2. $top\text{-}k_{CAM,high}$, $non\text{-}top\text{-}k_{CAM,high}$ \\
3. $top\text{-}k_{GradCAM,low}$, $non\text{-}top\text{-}k_{GradCAM,low}$ \\
4. $top\text{-}k_{GradCAM,high}$, $non\text{-}top\text{-}k_{GradCAM,high}$
}
\textbf{Initialize:} \
Create empty sets $S_{low}$ and $S_{high}$ for risk groups \\
\textbf{Step 1: Risk Classification} \\
\ForEach{window $w \in D$}{
Calculate ensemble prediction confidence interval $CI_w = [P_{25}, P_{75}]$ \\
\uIf{$P_{75} < threshold$}{
Add $w$ to $S_{low}$ \tcp*{Very low risk}}
\uElseIf{$P_{25} > threshold$}{
Add $w$ to $S_{high}$ \tcp*{Very high risk}}
\Else{
Discard window \tcp*{Uncertain risk}}}
\textbf{Step 2: Feature Importance Scoring} \\
\ForEach{XAI method $M \in {CAM, GradCAM}$}{
\ForEach{risk group $S \in {S_{low}, S_{high}}$}{
Initialize joint importance map $J_{M,S}$ \
\ForEach{window $w \in S$}{
\ForEach{joint $j$ in window}{
Calculate attribution scores using method $M$ \\
Compute confidence interval $CI_{score} = [P_{25}, P_{75}]$ of ensemble attribution scores \\
\uIf{$S = S_{low}$ \textbf{and} $P_{75} < t_M$}{
$J_{M,S}[j]$ += 1 \tcp*{Count very low scores with high confidence}
}
\uElseIf{$S = S_{high}$ \textbf{and} $P_{25} > t_M$}{
$J_{M,S}[j]$ += 1 \tcp*{Count very high scores with high confidence}}}}
Normalize $J_{M,S}$ by $|S|$ \tcp*{Divide by window count}}}
\textbf{Step 3: Determine Optimal k} \\
\ForEach{XAI method $M \in {CAM, GradCAM}$}{
\ForEach{importance map $J_{M,S}$}{
$k_1$ = FindKneePoint(SortDescending($J_{M,S}$)) \\
$k_2$ = KMeans($J_{M,S}$) \\
\ForEach{joint $j$}{
votes = CountSelections($j$, ${k_1, k_2}$) \\
\uIf{votes == $2$}{
Add $j$ to $top$-$k_{M,S}$}
\Else{
Add $j$ to $non$-$top$-$k_{M,S}$}}}}
\Return{All $top$-$k$ and $non$-$top$-$k$ sets}
\end{algorithm*}

\clearpage
\subsection{Velocity Perturbation}
\label{alg:velocity_perturbation}
\begin{algorithm*}[!ht]
\caption{Velocity Perturbation Analysis}
\KwIn{
    Test set $T$ containing 5-second skeletal windows \\
    Training/validation velocity percentiles $P5_{\text{ref}}$, $P95_{\text{ref}}$ \\
    Significant and non-significant joints from XAI methods (CAM, Grad-CAM) \\
    Predefined segments $S$ of biomechanically linked joints \\
    Slowdown factors $F_{\text{slow}} = [0.20, 0.25, 0.33, 0.5, 1]$ \\
    Speedup factors $F_{\text{fast}} = [1, 2, 3, 4, 5]$
}
\KwOut{
    Perturbed windows and their predicted risk scores
}
\BlankLine
\For{each window $w \in T$}{
    \tcc{Step 1: Calculate sample-specific scaling}
    Compute $P5_{\text{sample}}$, $P95_{\text{sample}}$ for each joint $j$ \\
    $s_{\text{min}}(j) = P5_{\text{ref}}(j) / P5_{\text{sample}}(j)$ \\
    $s_{\text{max}}(j) = P95_{\text{ref}}(j) / P95_{\text{sample}}(j)$
    
    \tcc{Step 2: Apply biomechanical constraints}
    \For{each segment $s \in S$}{
        $s_{\text{min}}(s) = \max\{s_{\text{min}}(j) | j \in s, s_{\text{min}}(j) \leq 1\}$ \\
        $s_{\text{max}}(s) = \min\{s_{\text{max}}(j) | j \in s, s_{\text{max}}(j) \geq 1\}$
    }
    
    \tcc{Step 3: Generate and evaluate perturbations}
    \For{each factor $f \in F_{\text{slow}}$}{
        $w_{\text{perturbed}} = \text{InterpolateTrajectories}(w, f \times s_{\text{min}})$ \\
        Record model prediction for $w_{\text{perturbed}}$
    }
    \For{each factor $f \in F_{\text{fast}}$}{
        $w_{\text{perturbed}} = \text{InterpolateTrajectories}(w, f \times s_{\text{max}})$ \\
        Record model prediction for $w_{\text{perturbed}}$
    }
}
\Return{All perturbed window predictions for both CAM and Grad-CAM analyses}
\end{algorithm*}

\clearpage
\subsection{Angular Perturbation}
\label{alg:angle_perturbation}
\begin{algorithm*}[!ht]
\caption{Angular Perturbation Analysis}
\KwIn{
    Test set $T$ containing 5-second skeletal windows \\
    Training/validation angular change percentiles $P5_{\text{ref}}$, $P95_{\text{ref}}$ \\
    Significant and non-significant joints from XAI methods (CAM, Grad-CAM) \\
    Joint adjacency map $A$ defining parent-child joint relationships \\
    Reduction factors $F_{\text{reduce}} = [0.20, 0.25, 0.33, 0.5, 1]$ \\
    Amplification factors $F_{\text{amplify}} = [1, 2, 3, 4, 5]$
}
\KwOut{
    Perturbed windows and their predicted risk scores
}
\BlankLine
\For{each window $w \in T$}{
    \tcc{Step 1: Calculate sample-specific scaling}
    Compute $P5_{\text{sample}}$, $P95_{\text{sample}}$ for each joint $j$ \\
    $s_{\text{min}}(j) = P5_{\text{ref}}(j) / P5_{\text{sample}}(j)$ \\
    $s_{\text{max}}(j) = P95_{\text{ref}}(j) / P95_{\text{sample}}(j)$
    
    \tcc{Step 2: Process each frame}
    \For{frame $t = 1$ to total frames $n$}{
        \For{each joint $j$ to perturb}{
            parent = $A[j]$ \\
            $\theta_{\text{curr}} = \arctan2(y_j^t - y_{\text{parent}}^t, x_j^t - x_{\text{parent}}^t)$ \\
            $\theta_{\text{prev}} = \arctan2(y_j^{t-1} - y_{\text{parent}}^{t-1}, x_j^{t-1} - x_{\text{parent}}^{t-1})$ \\
            $\Delta = \theta_{\text{curr}} - \theta_{\text{prev}}$
            
            \tcc{Step 3: Apply perturbations}
            \For{each factor $f \in F_{\text{reduce}}$}{
                $\Delta_{\text{new}} = f \times s_{\text{min}}(j) \times \Delta$ \\
                Update joint position $(x_j^t, y_j^t)$ using $\Delta_{\text{new}}$ while preserving limb length
            }
            \For{each factor $f \in F_{\text{amplify}}$}{
                $\Delta_{\text{new}} = f \times s_{\text{max}}(j) \times \Delta$ \\
                Update joint position $(x_j^t, y_j^t)$ using $\Delta_{\text{new}}$ while preserving limb length
            }
        }
    }
}
\Return{All perturbed window predictions for both CAM and Grad-CAM analyses}
\end{algorithm*}

\end{document}